\documentstyle [12pt]{article}
\title{Conditional Lie-B\"acklund symmetry and reduction
of evolution equations.}
\author{R.Z.Zhdanov \\ \small Arnold-Sommerfeld Institute for Mathematical
Physics,\cr \small Leibnitzstra\ss e 10, 38678 Clausthal-Zellerfeld, Germany
\thanks {On leave from the Institute of Mathematics of the Academy of
Sciences of Ukraine, Tereshchenkivska Str.3, 252004 Kiev, Ukraine
\newline\indent e-mail: asrz@pta3.pt.tu-clausthal.de}}
\input amssym.def
\let \ds \displaystyle
\begin{document}
\maketitle
\begin{abstract} We suggest a generalization of the notion of
  invariance of a given partial differential equation with respect to
  Lie-B\"acklund vector field. Such generalization proves to be
  effective and enables us to construct principally new Ans\"atze
  reducing evolution-type equations to several ordinary differential
  equations. In the framework of the said generalization we obtain
  principally new reductions of a number of nonlinear heat
  conductivity equations $u_t=u_{xx}+F(u,u_x)$ with poor Lie symmetry
  and obtain their exact solutions. It is shown that these solutions
  can not be constructed by means of the symmetry reduction procedure.
\end{abstract}

\section{Introduction}
Construction of exact solutions of nonlinear partial differential
equations (PDEs) is one of the most important problems of the modern
mathematical physics. The most effective and universal method used is
the symmetry reduction procedure pioneered by Sophus Lie. But there is
a natural restriction on the application of the said procedure:
equation under study should have non-trivial Lie symmetry.  There
exist very important equations (in particular, the ones describing
heat conductivity and some nonlinear processes in biology) with very
poor Lie symmetry. So it would be desirable to modify the symmetry
reduction procedure in such a way that it could be applied to these
equations as well. Fortunately, the main idea of the symmetry
reduction procedure -- the reduction of the equation under study to
PDEs having less number of independent variables by means of specially
chosen Ans\"atze -- can be applied to some of these if one utilizes
their {\it conditional} symmetry (see \cite{FuN}, \cite{FuZ1}). The
method of conditional symmetries of PDEs is closely connected with the
``non-classical reduction'' \cite{BlC} and ``direct reduction''
\cite{ClKr} methods (see also \cite{LeW}, \cite{OlR}).

Further possibility of constructing exact solutions of PDEs is to use
their Lie-B\"acklund (higher, generalized) symmetry \cite{Ib}. In this
way multi-soliton solutions of the KdV, mKdV, sine-Gordon and cubic
Schr\"odinger equations can be obtained \cite{FuC}. But the choice of
physically significant examples of equations admitting non-trivial
Lie-B\"acklund symmetry is very restricted. On the other hand, there
are examples due to Galaktionov \cite{Ga,Sa} and Fushchych et al
\cite{FuM, FuR} of Ans\"atze reducing PDEs admitting only trivial
Lie-B\"acklund symmetry to systems of ordinary differential equations
(ODEs). This facts can be understood within the framework of the {\it
  conditional} Lie-B\"acklund symmetry which is introduced below. It
will be established that conditional invariance of the equation under
study ensures its reducibility and can be applied to construct its
exact solutions. Since the class of PDEs conditionally-invariant with
respect to some Lie-B\"acklund field is substantially wider than the
one of PDEs admitting Lie-B\"acklund symmetry in the classical sense,
the said result yields principally new possibilities of reduction of
PDEs with poor Lie and Lie-B\"acklund symmetry. We will give several
examples of reduction of PDEs to systems of ODEs by means of the
An\"atze corresponding to their conditional Lie-B\"acklund symmetry
and will show that the exact solutions obtained in this way can not be
constructed by means of classical symmetry reduction procedure.

Let

\begin{equation}
  \label{1}
  u_t=F(t,\, x,\, u,\, u_1,\, u_2,\ldots ,u_n),
\end{equation}
where $u\in {C^n(\Bbb{R}^2,\Bbb{C}^1)}, \ u_k=\partial ^ku/\partial
x^k, \ 1\le k \le n$, be some evolution-type equation and

\begin{equation}
  \label{2}
  Q=\eta \partial_u+(D_x\eta )\partial_{u_1}+(D_t\eta)\partial_{u_t}+
  (D_x^2\eta )\partial_{u_2}+\ldots
\end{equation}
with
\begin{equation}
  \label{3}
  \eta =\eta (t,\, x,\, u,\, u_t,\, u_1,\, u_{tt},\, u_{t1},\ldots )
\end{equation}
be some smooth Lie-B\" acklund vector field (LBVF).

In the above formulae we denote by the symbols $D_t$ and $D_x$ the
total differentiation operators  with respect to the variables $t$ and
$x$ correspondingly, i.e.

\begin{eqnarray*}
 D_t&=&\partial_t+u_t\partial_u+u_{tt}\partial{u_t}+
 u_{t1}\partial{u_1}+\ldots,\\
 D_x&=&\partial_x+u_1\partial_u+u_{t1}\partial{u_t}+
 u_2\partial{u_1}+\ldots.
\end{eqnarray*}
\vskip 1.5mm

If the function $\eta$ is of the form
\begin{equation}
  \label{3*}
  \eta = \tilde\eta(t,x,u)-\xi_0(t,x,u)u_t-\xi_1(t,x,u)u_x,
\end{equation}
then the LBVF (\ref{2}) is equivalent to the usual Lie vector field
and can be represented in the following equivalent form:
\begin{displaymath}
  Q=\xi_0(t,x,u)\partial_t+\xi_1(t,x,u)\partial_x+
  \tilde\eta(t,x,u)\partial_u.
\end{displaymath}
{\bf Definition 1.}\ We say that Eq.(\ref{1}) is invariant under the
LBVF (\ref{2}) if the condition

\begin{equation}
  \label{4}
  \left . \matrix {Q(u_t-F)\cr \cr }\right.  \left| \matrix{=0\cr
  {\scriptstyle M} \cr }\right.
\end{equation}
holds.

In (\ref{3}) $M$ is a set of all differential consequences of
the equation $u_t-F=0$.
\vskip 1.5mm

{\bf Definition 2.}\ We say that Eq.(\ref{1}) is
conditionally-invariant under LBVF (\ref{2}) if the following
condition

\begin{equation}
  \label{5}
  \left . \matrix {Q(u_t-F)\cr \cr }\right.  \left|
  \matrix{=0\cr {\scriptstyle M\cap L_{x}}\cr }\right.
\end{equation}
holds.

Here the symbol $L_{x}$ denotes the set of all differential
consequences of the equation $\eta =0$ with respect to the variable
$x$.

Evidently, condition (\ref{3}) is nothing else than a usual invariance
criteria for Eq.(\ref{1}) under LBVF (\ref{2}) written in a canonical
form (see, e.g. \cite{Ib}). The most of ``soliton equations'' like the
KdV, mKdV, cubic Schr\"odinger, sine-Gordon equations admit infinitely
many LBVFs which can be obtained from some initial LBVF by applying
the recursion operator.

Another important remark is that on the set of solutions of
Eq.(\ref{1}) we can exclude all derivatives with respect to $t$ and
thus get the vector field (\ref{2}) with $\eta $ of the form

\begin{equation}
  \label{6}
  \eta =\eta (t,x,u,u_1,u_2,\ldots,u_N).
\end{equation}

In the following we will consider LBVFs of the form (\ref{2}),
(\ref{6}) only.

Clearly, if Eq.(\ref{1}) is invariant under LBVF (\ref{2}), then it is
conditionally-invariant under the said field. But the inverse
assertion is not true. This means, in particular, that the Definition
2 is a generalization of the standard definition of invariance of
partial differential equation with respect to LBVF. Provided
(\ref{2}) is a Lie vector field, Definition 2 coincides with the one
of $Q$-conditional invariance under the Lie vector field.

One of the important consequences of  $Q$-conditional invariance of
a given PDE under the Lie vector field is a possibility to get an
Ansatz which reduces this PDE to one PDE with less number of
independent variables (see, e.g. \cite{FuZ1}). We will show that
conditional invariance of the evolution-type equation (\ref{1})
ensures its reducibility to $N$ ordinary differential equations (ODEs)
($N$ is the order of the highest derivative contained in $\eta $ from
(\ref{6})).

\section{Reduction theorem}
Consider the nonlinear PDE
\begin{equation}
  \label{7}
  \eta (t,\, x,\, u,\, u_1,\ldots,u_N)=0
\end{equation}
as the $N$-th order ODE with respect to variable $x$.
Its general integral is written (at least locally) in the form

\begin{equation}
  \label{8}
  u=f\Bigl (t,\, x,\, \varphi_1(t),\,  \varphi_2(t),\ldots
  ,\varphi_N(t)\Bigr ),
\end{equation}
where $\varphi_j(t), \ j=\overline {1,N}$ are arbitrary smooth
functions. We will call the expression (\ref{8}) an Ansatz invariant
under LBVF (\ref{2}), (\ref{6}).
\vskip 2mm
{\bf Theorem 1.}\ {\it Let Eq.(\ref{1}) be conditionally-invariant
  under the LBVF (\ref{2}), (\ref{6}). Then the Ansatz (\ref{8})
  invariant under LBVF (\ref{2}), (\ref{6}) reduces PDE (\ref{1}) to a
  system of $N$ ODEs for functions $\varphi_j(t), \
  j=\overline{1,N}$.}

{\sl P r o o f}.\ We first proof that given the conditions of the
theorem the system of PDEs

\begin{equation}
  \label{9}
  \cases {u_t=F(t,\, x,\, u,\, u_1,\ldots ,u_n),\cr \cr
\eta (t,\, x,\, u,\, u_1,\ldots ,u_N)=0\cr }
\end{equation}
is compatible.

Differentiating the first equation from (\ref{9}) $N$ times with
respect to $x$, the second -- one time with respect to $t$ and
comparing the derivatives $u_{N t}$ and $u_{t N}$ we get the equality

$$D_x^NF=-(\eta _{u_N})^{-1}(\eta _t+\eta _uu_t+\eta
_{u_1}u_{1t}+\ldots +\eta _{u_{N-1}}u_{N-1 t})$$
or

$$D_x^NF=-(\eta _{u_N})^{-1}(\eta _t+\eta _uF+\eta _{u_1}D_xF+
\ldots +\eta _{u_{N-1}}D_x^{N-1}F).$$

Consequently, provided the condition

\begin{equation}
  \label{10}
  \left . \matrix {(\eta _t+\eta _uF+\eta _{u_1}D_xF+\ldots +\eta
  _{u_N}D_x^NF)\cr \cr }\right.  \left| \matrix{=0\cr
  {\scriptstyle M\cap L}\cr }\right.
\end{equation}
where $L$ is the set of all differential consequences of the equation
$\eta=0$, holds identically, the system of PDEs (\ref{9}) is in
involution and its general solution contains $N$ arbitrary complex
constants $C_1,\ C_2,\ldots ,C_N$ \cite{Po}.

We will prove that the relation (\ref{10}) follows from (\ref{5}).

Really, with account of (\ref{2}) the equality (\ref{5}) is rewritten
in the form
$$\left . \matrix {D_t\eta -\eta F_u- (D_x\eta )F_{u_1}-\ldots -
(D_x^n\eta )F_{u_n}\cr \cr }\right.  \left| \matrix{=0\cr
{\scriptstyle M\cap L_{x}}\cr }\right. $$
or

$$\left . \matrix {D_t\eta \cr \cr }\right.
\left|\matrix{=0.\cr {\scriptstyle M\cap L_{x}}\cr }\right. $$

Since $D_t\eta =\eta _t+\eta _uu_t+\eta _{u_1}u_{1t}+
\ldots +\eta_{u_N}u_{N t}$, the above equation reads

$$\left . \matrix {\eta _t+\eta _uu_t+\eta _{u_1}u_{1t}+\ldots +
\eta _{u_N}u_{N t}  \cr \cr }\right.
\left |\matrix {=0,\cr {\scriptstyle M\cap L_{x}}\cr }\right. $$
whence

\begin{equation}
  \label{11}
  \left . \matrix {\eta _t+\eta _uF+\eta _{u_1}D_xF+\ldots +\eta
      _{u_N}D_x^NF \cr \cr }\right.  \left|\matrix{=0.\cr
      {\scriptstyle M\cap L_{x}} \cr }\right.
\end{equation}

Since the manifold $M\cap L$ is contained in the manifold $M\cap L_{x}$,
relation (\ref{10}) follows from relation (\ref{11}).

Next, we consider the determinant
\begin{equation}
  \label{12}
  \Delta =\left |\matrix {
  \ds {\partial f\over \partial \varphi _1} & \ds {\partial f\over
    \partial \varphi _2} &
  \cdots & \ds {\partial f\over \partial \varphi _N}\cr
  \ds {\partial ^2 f\over \partial \varphi _1\partial x} &
  \ds {\partial ^2 f\over \partial \varphi _2\partial x} & \cdots &
  \ds {\partial ^2 f\over \partial \varphi _N\partial x}\cr
  \vdots & \vdots & \ddots & \vdots \cr
  \ds {\partial ^N f\over \partial \varphi _1\partial {x^{N-1}}} &
  \ds {\partial ^N f\over \partial \varphi _2\partial {x^{N-1}}} & \cdots &
  \ds {\partial ^N f\over \partial \varphi _N\partial {x^{N-1}}}\cr }\right|
\end{equation}

The above determinant $\Delta$ is the Wronsky determinant for
functions $y_j={\partial f\over \partial \varphi _j}, \ j = \overline
{1,N}$. We will prove that in the case involved $\Delta \not= 0$.

Let $\Delta =0$, then due to the properties of the Wronsky
determinant the functions $y_j$ are linearly-dependent. Consequently,
there exist such $\lambda _j =\lambda _j(t), \ j = \overline {1,N}$ that

$$\sum_{j=1}^N \lambda _j(t)y_j=0. $$

Substituting into the above equality $y_j={\partial f\over \partial
\varphi _j}$ we get

\begin{equation}
  \label{13}
  \sum_{j=1}^N \lambda _j(t){\partial f\over \partial \varphi _j}=0.
\end{equation}

Integrating the first-order PDE (\ref{13}) we have

$$f=\tilde f(t,\, x,\, \omega _1,\, \omega _2,\ldots ,\omega
_{N-1}),$$
where $\omega _j=\lambda _N\varphi _j-\lambda _j\varphi _N,
\ j = \overline {1,N-1}$.

Consequently, in the case $\Delta =0$ the general solution of the ODE
(\ref{7}) depends not on $N$ but on $N-1$ arbitrary constants $\omega
_j(t), \ j = \overline {1,N-1}$. We arrive at the contradiction, a
source of which is an assumption that $\Delta =0$. Hence we conclude
that $\Delta \not= 0$.

Substituting (\ref{8}) into (\ref{1}) we obtain

$$\sum_{j=1}^N \dot \varphi _j{\partial f\over \partial \varphi _j}=
-f_t+F(t,\, x,\, {\partial f\over \partial x},\, {\partial ^2 f\over
  \partial x^2},\ldots , {\partial ^nf\over \partial x^n})$$
or

\begin{equation}
  \label{14}
  \sum_{j=1}^N \dot \varphi _j{\partial f\over \partial \varphi _j}=
  G\Bigl (t,\, x,\, \varphi_1(t),\, \varphi_2(t),\ldots ,\varphi
  _N(t)\Bigr ).
\end{equation}

Hereafter, an overdot means differentiation with respect to $t$.

Differentiation of (\ref{14}) $N-1$ times with respect to the variable
$x$ yields the following result:

\begin{equation}
  \label{15}
  \sum_{j=1}^N \dot \varphi _j{\partial ^{k+1}f\over
  \partial \varphi _j\partial x^k}=
  {\partial ^kG\over \partial x^k}, \quad k=\overline {1,N-1}.
\end{equation}

If we consider Eqs.(\ref{14}), (\ref{15}) as a system of linear
inhomogeneous algebraic equations for functions $\dot \varphi _1,\
\dot \varphi _2, \ldots ,\dot \varphi _N$, then its determinant has
the form (\ref{12}) and, consequently, is not equal to zero. Solving
(\ref{14}), (\ref{15}) with respect to the functions $\dot \varphi _j,
\ j=\overline {1,N}$ we get

\begin{equation}
  \label{16}
  \dot \varphi _j=H_j(t,\, x,\, \varphi _1,\, \varphi _2,\ldots
  ,\varphi_N), \quad j=\overline {1,N}.
\end{equation}

Let us expand the right-hand sides of (\ref{16}) into a Taylor series
with respect to the variable $x$ in the neighbourhood of $x_0$ and
then equate coefficients at $(x-x_0)^k$

\begin{equation}
  \label{17a}
  \dot \varphi _j=H_j(t,\, x_0,\, \varphi _1,\, \varphi _2,\ldots
  ,\varphi_N), \quad j=\overline {1,N},
\end{equation}

\begin{equation}
  \label{17b}
  0={\partial ^k H_j\over \partial x^k}(t,\, x_0,\, \varphi _1,\,
  \varphi _2,\ldots ,\varphi _N),\quad j=\overline {1,N},\quad k \ge
  1.
\end{equation}

Thus, we have established that the system of PDEs (\ref{9}) is
equivalent to the infinite set of Eqs.(\ref{17a}), (\ref{17b}).

Next, we will prove that right-hand sides of Eqs.(\ref{17b}) vanish
identically on the solutions of the system of ODEs (\ref{17a}).

Let $\varphi_j= \tilde \varphi_j(t,\, C_1,\, C_2,\ldots ,C_N), \ j =
\overline {1,N}$ where $C_j$ are arbitrary complex constants, be a
general solution of the system of ODEs (\ref{17a}). If at least one of
the equations is not satisfied identically on the solutions of
Eqs.(\ref{17a}), then substituting into it the expressions for
$\varphi _j$ we get a relation of the form $h(C_1,C_2,\ldots ,C_N)=0$.
Hence it follows that the general solution of the system of PDEs
(\ref{9}) contains no more than $N-1$ independent constants. We arrive
at the contradiction, which proves that the right-hand sides of
Eqs.(\ref{17b}) vanish identically on the solutions of system of ODEs
(\ref{17a}). Consequently, system (\ref{17a}), (\ref{17b}) is
equivalent to system of $N$ ODEs

\begin{equation}
  \label{18}
  \dot \varphi _j=H_j(t,\, x_0,\, \varphi _1,\, \varphi _2,\ldots
  ,\varphi_N)= \tilde H_j(t,\, x,\, \varphi _1,\, \varphi _2,\ldots
  ,\varphi_N), \quad j=\overline {1,N}.
\end{equation}

Thus, given the conditions of the theorem the Ansatz (\ref{8}), which
is invariant under LBVF (\ref{2}), (\ref{6}), reduces the equation
(\ref{1}) to system of $N$ ODEs (\ref{18}). The theorem is proved.
\vskip 1.5mm

{\bf Consequence.} {\it Let Eq.(\ref{1}) be invariant under the LBVF
(\ref{2}), (\ref{6}). Then the Ansatz (\ref{8}) invariant under LBVF
(\ref{2}), (\ref{6}) reduces PDE (\ref{1}) to a system of $N$ ODEs for
functions $\varphi_j(t), \ j=\overline{1,N}$.}

Proof follows from the fact that if an equation is invariant
under LBVF, then it is conditionally-invariant with respect
to this LBVF.

\section{Some examples}
Utilizing the above theorem one can construct principally
new exact solutions even for equations with poor Lie
symmetry. As an illustration, we give below several examples.
\vskip 1.5mm

\noindent
{\it Example 1.} Consider the nonlinear heat conductivity equation with
a logarithmic-type nonlinearity

\begin{equation}
  \label{19}
  u_t=u _{xx}+ \Bigl (\alpha +\beta \ln u - \gamma ^2 (\ln u )^2\Bigr
  ) u.
\end{equation}

We will establish that Eq.(\ref{19}) is conditionally-invariant
with respect to LBVF (\ref{2}) with

\begin{equation}
  \label{20}
  \eta = u_ {xx} - \gamma u_x-u^{-1}u_x^2.
\end{equation}

Condition (\ref{5}) for Eq.(\ref{19}) reads

\begin{equation}
  \label{21}
  \left . \matrix {D_t\eta -D_x^2\eta -\Bigl (\alpha +\beta +
(\beta -2\gamma^2)\ln u-\gamma ^2\ln ^2u\Bigr )\eta \cr \cr }
\right.  \left|\matrix{=0,\cr {\scriptstyle M\cap L_{x}}\cr
  }\right.
\end{equation}
where $M$ stands for the set of all differential consequences of the
equation $u_t=u _{xx}+ \Bigl (\alpha +\beta \ln u - \gamma ^2
(\ln u)^2\Bigr )u$ and $L_{x}$ stands for the set of all differential
consequences of the equation $u_ {xx} - \gamma u_x-u^{-1}u_x^2=0$ with
respect to $x$.

Substituting into the left-hand side of Eq.(\ref{21}) expression
(\ref{20}) and transferring to the manifold $M$ (i.e. excluding the
derivatives $u_t$, $u_{tx}$, $u_{txx}$ with the help of Eq.(\ref{19}))
we transform it to the form

$$2u^{-1}(u_ {xx} - \gamma u_x-u^{-1}u_x^2)^2+  4\gamma
u^{-1}u_x(u_ {xx} - \gamma u_x-u^{-1}u_x^2).$$

Evidently, the above expression does not vanish on the manifold $M$.
But on the manifold $M\cap L_{x}$ it vanishes identically

$$\left . \matrix {2u^{-1}(u_ {xx} - \gamma u_x-u^{-1}u_x^2)^2+
  4\gamma u^{-1}u_x(u_ {xx} - \gamma u_x-u^{-1}u_x^2) \cr \cr }\right.
\left|\matrix{=0.\cr {\scriptstyle M\cap L_{x}}\cr }\right. $$

Hence it follows that the nonlinear heat conductivity equation
(\ref{19}) is conditionally-invariant under LBVF (\ref{2}) with $\eta$
of the form (\ref{20}) but not invariant under the said LBVF in the
sense of Definition 1. It is also seen from \cite{Ib}, where the
results of classification of nonlinear heat conductivity equations
$u_t=u_{xx}+F(u)$ admitting LBVF are given. It has been established, in
particular, that only linear heat equation admits LBVF which can not
represented in the form (\ref{2}), (\ref{3*}) and, consequently, is
not equivalent to a Lie vector field.

Integrating the equation $\eta \equiv u_ {xx} - \gamma
u_x-u^{-1}u_x^2=0$ as an ODE with respect to $x$ we get
an Ansatz for $u(t,x)$

\begin{equation}
  \label{22}
  u(t,x)=\exp\Bigl(\varphi _1(t)+\varphi _2(t)\exp {\gamma x}\Bigr).
\end{equation}

Substitution of the Ansatz (\ref{22}) into Eq.(\ref{19}) gives rise to  a
system of two ODEs

$$\dot \varphi _1 = \alpha + \beta\varphi _1-\gamma ^2\varphi _1^2,
\quad \dot\varphi _2=(\beta +\gamma^2-2\gamma ^2\varphi _1)\varphi _2.$$

The general solution of the above system is given by one of the
following formulae:

(a) $k =\beta ^2+4\alpha \gamma ^2>0$

$$u=C\biggl (\cos {{k ^{1/2}t\over 2}}\biggr )^2\exp (\gamma x
+\gamma ^2t)+{1\over 2\gamma ^2}\biggl (\beta -k ^{1/2}\
\tan{k ^{1/2}t\over 2}\biggr );$$

(b) $k =\beta ^2+4\alpha \gamma ^2<0$

$$u=C\biggl (\cosh {{(-k)^{1/2}t\over 2}}\biggr )^2\exp (\gamma x
+\gamma ^2t)+{1\over 2\gamma ^2}\biggl (\beta +(-k ) ^{1/2}
\tanh {(-k) ^{1/2}t\over 2}\biggr );$$

(c) $k =\beta ^2+4\alpha \gamma ^2=0$

$$u=Ct^{-2}\exp(\gamma x+\gamma ^2t)+{1\over2\gamma ^2t}(\beta t+2).$$

Here $C$ is an arbitrary constant.

It is important to emphasize that the above solutions can not be
obtained by the symmetry reduction procedure. Really, the maximal
local invariance group of Eq.(\ref{19}) is the two-parameter group of
translations \cite{Ov}

$$t^\prime =t+\theta _1,\ x^\prime =x+\theta _2, \ u^\prime  =u.$$
and solutions (a)--(b) are obviously not invariant under the
above group.
\vskip 1.5mm

\noindent
{\it Example 2.} Consider the following nonlinear heat conductivity
equation:

\begin{equation}
  \label{23}
  u_t=u _{xx}+F(u).
\end{equation}

We will establish that it is conditionally-invariant
with respect to the LBVF (\ref{2}) with $\eta =u_{xx}
-A(u)u_x^2$, provided functions $F(u)$, \ $A(u)$ satisfy the
following system of ODEs:

\begin{equation}
  \label{24}
  \ddot A+4A\dot A+2A^3=0,\quad \ddot F-\dot AF-A\dot F=0.
\end{equation}

The equality (\ref{5}) for Eq.(\ref{23}) takes the form

$$\left . \matrix {D_t\eta -D_x^2\eta -\dot F\eta \cr \cr }
\right.  \left|\matrix{=0,\cr {\scriptstyle M\cap L_{x}}\cr
  }\right. $$
where $M$ is the set of all differential consequences of the equation
$u_t=u _{xx}+ F(u)$ and $L_{x}$ is the set of all differential
consequences of the equation $u_ {xx} - A(u)u_x^2=0$ with respect to
$x$.

Excluding from the left-hand side of the above equality the derivatives
$u_t$, \ $u_{tx}$, \ $u_{txx}$ and grouping in a proper way terms in
the expression obtained we have

$$\left . \matrix {2A\eta ^2+4(\dot A+A^2)\eta +
(\ddot A+4A\dot A+2A^3)u_x^4+(\ddot F-\dot AF-A\dot F)u_x^2\cr \cr }
\right.  \left|\matrix{=0\cr {\scriptstyle M\cap L_{x}}\cr }\right. $$
or with account of Eqs.(\ref{24})

\begin{equation}
  \label{25}
  \left . \matrix {2A\eta ^2+4(\dot A+A^2)\eta \cr \cr }\right.
  \left|\matrix{=0.\cr {\scriptstyle M\cap L_{x}}\cr
    }\right.
\end{equation}

Evidently, the left-hand side of Eq.(\ref{25}) does not vanish on the
manifold $M$ but it does vanish on the manifold $M\cap L_x$.
Consequently, the nonlinear heat equation (\ref{23}) is
conditionally-invariant with respect to LBVF (\ref{2}) with $\eta
=u_{xx}-A(u)u_x^2$ iff Eqs.(\ref{24}) hold.

Thus, conditions of the Theorem 1 are satisfied and we can reduce
Eq.(\ref{23}) to two ODEs with the help of the Ansatz (\ref{8})
invariant under the above mentioned LBVF.

Let the function $\theta (u)$ be determined by the following
equality:

$$\int \limits _{\ds 0}^{\ds\theta(u)}(\ln \tau)^{-1/2}d\tau =
\alpha u+\beta, $$
where $\alpha $,\ $\beta $ are arbitrary real constants. Then the
Ansatz

$$\int \limits _{\ds 0}^{\ds u(t,x)}{d\tau \over \theta (\tau )}=
x\varphi _1(t)+\varphi _2(t)$$
reduce the nonlinear equation (\ref{23}) with

\begin{equation}
\label{26}
F(u)=\Biggl (\lambda _1+\lambda _2\int \limits _{\ds 0}^{\ds u}
{d\tau \over \theta (\tau )}\Biggr )\theta (u)
\end{equation}
to the following system of ODEs:

$$\dot \varphi _1=\biggl ({\alpha ^2\over 2}\varphi _1^2+\lambda
_2\biggr )\varphi _1,
\quad \dot \varphi _2=\biggl ({\alpha ^2\over 2}\varphi _1^2+\lambda
_2\biggr )\varphi _2+
\dot \theta (0)\varphi _1^2+\lambda _1.$$

Here $\lambda _1$,\ $\lambda _2$ are arbitrary real constants
and $\dot \theta (0)$ is a value of the first derivative of
the function $\dot\theta (u)$ in the point $x=0$.

The above system of ODEs is integrated in quadratures thus giving
rise to a family of exact solutions of the nonlinear PDE (\ref{23})
with rather exotic nonlinearity (\ref{26}). The solutions obtained
are also non-invariant with respect to the two-parameter
group of translations with respect to $t$, $x$, which is
the maximal local invariance group of Eq.(\ref{23}), (\ref{26}).
\vskip 1.5mm

\noindent
{\it Example 3.}\
Here, we will perform the reduction of a nonlinear PDE of the form
(\ref{19})
\begin{equation}
  \label{27}
  u_t=u_{xx}+a(\ln^2u)u,\ \ a\in \Bbb{R}^1
\end{equation}
to systems of three ODEs.

By a rather cumbersome computation one can check that Eq.(\ref{27}) is
conditionally-inva\-ri\-ant with respect to the LBVF (\ref{2}) with
\begin{equation}
  \label{28}
  \eta=u^2u_{xxx}-3uu_xu_{xx}+2u_x^3+au_xu^2.
\end{equation}

Integrating the third-order ODE $\eta=0$ we obtain the following
Ans\"atze for the function $u(t,x)$:
\vskip 1.5mm

\noindent
1)\ under $a=\alpha^2>0$
\begin{equation}
  \label{29}
  u(t,x)=\exp\{\varphi_1(t)+\varphi_2(t)\cos\alpha x
  +\varphi_3(t)\sin\alpha x\},
\end{equation}
2)\ under $a=-\alpha^2<0$
\begin{equation}
  \label{30}
  u(t,x)=\exp\{\varphi_1(t)+\varphi_2(t)\cosh\alpha x
  +\varphi_3(t)\sinh\alpha x\},
\end{equation}
where $\varphi_1,\ \varphi_2,\ \varphi_3$ are arbitrary smooth
functions.

Substitution of the expressions (\ref{29}), (\ref{30}) into PDE
(\ref{27}) gives rise to the systems of nonlinear ODEs for the
functions $\varphi_1,\ \varphi_2,\ \varphi_3$
\vskip 1.5mm

\noindent
1)\ under $a=\alpha^2>0$
\begin{eqnarray*}
  \dot \varphi_1&=&\alpha^2(\varphi_1^2 + \varphi_2^2 +
  \varphi_3^2),\\
  \dot \varphi_2&=&\alpha^2(2\varphi_1 - 1)\varphi_2,\\
  \dot \varphi_3&=&\alpha^2(2\varphi_1 - 1)\varphi_3;
\end{eqnarray*}
2)\ under $a=-\alpha^2<0$
\begin{eqnarray*}
  \dot \varphi_1&=&\alpha^2(\varphi_3^2 - \varphi_2^2 -
  \varphi_1^2),\\
  \dot \varphi_2&=&\alpha^2(1 - 2\varphi_1)\varphi_2,\\
  \dot \varphi_3&=&\alpha^2(1 - 2\varphi_1)\varphi_3.
\end{eqnarray*}

Making the change of the dependent variable $u = \exp v$ we rewrite
Eq.(\ref{27}) in the form
\begin{equation}
  \label{31}
  v_t = v_{xx} + v_x^2+ av^2
\end{equation}
and what is more, the Ans\"atze (\ref{29}), (\ref{30}) take the form
\vskip 1.5mm

\noindent
1)\ under $a=\alpha^2>0$
\begin{equation}
  \label{32}
  v(t,x)=\varphi_1(t)+\varphi_2(t)\cos\alpha x
  +\varphi_3(t)\sin\alpha x,
\end{equation}
2)\ under $a=-\alpha^2<0$
\begin{equation}
  \label{33}
  v(t,x)=\varphi_1(t)+\varphi_2(t)\cosh\alpha x
  +\varphi_3(t)\sinh\alpha x.
\end{equation}

If we choose in formulae (\ref{32}), (\ref{33}) $\varphi_3=0$, then
the well-known Galaktionov's Ans\"atze are obtained \cite{Ga,Sa}. These
Ans\"atze were used to study blow-up solutions of the nonlinear PDE
(\ref{31}). It should be said that all solutions of the nonlinear heat
conductivity equations obtained in \cite{Ga} can be constructed within
the framework of our approach.
\vskip 1.5mm

\noindent
{\it Example 4.}\ Let us describe all PDEs of the form
\begin{equation}
  \label{34}
  u_t=u_{xx}+R(u, u_x)
\end{equation}
which are conditionally-invariant under the LBVF (\ref{2}) with
$\eta=u_{xx} - au,\ a\in\Bbb{R}^1$.

Acting by the operator (\ref{2})
on the equation (\ref{34}) and transferring to the manifold $M\cap
L_x$ we obtain the determining equation for the function $R$
\begin{displaymath}
  a^2u^2R_{u_xu_x} + 2auu_xR_{uu_x} + u_x^2R_{uu} + auR_u + au_xR_{u_x}
  + aR=0.
\end{displaymath}

The above PDE is rewritten in the form
\begin{displaymath}
  (J^2 + a)R=0,
\end{displaymath}
where $J = u_x\partial_u + au\partial_{u_x}$. With this remark it is
easily integrated and its general solution reads
\begin{displaymath}
  R = f_1(u_x^2 - au^2)u_x + f_2(u_x^2 - au^2)u.
\end{displaymath}

Here $f_1,\ f_2$ are arbitrary smooth functions.

Thus, the most general PDE of the form (\ref{34})
conditionally-invariant with respect to LBVF (\ref{2}) with
$\eta=u_{xx} - au$ is as follows
\begin{equation}
  \label{35}
  u_t=u_{xx} + f_1(u_x^2 - au^2)u_x + f_2(u_x^2 - au^2)u.
\end{equation}

Solving the equation $\eta\equiv u_{xx} - au = 0$ we obtain the
Ans\"atze for $u(t,x)$
\vskip 1.5mm

\noindent
1)\ under $a=-\alpha^2<0$
\begin{displaymath}
  u(t,x)=\varphi_1(t)\cos\alpha x
  +\varphi_2(t)\sin\alpha x,
\end{displaymath}
2)\ under $a=\alpha^2>0$
\begin{displaymath}
  u(t,x)=\varphi_1(t)\cosh\alpha x
  +\varphi_2(t)\sinh\alpha x
\end{displaymath}
which reduce PDE (\ref{35}) to systems of two ODEs for functions
$\varphi_1(t),\ \varphi_2(t)$
\begin{eqnarray*}
  &1)&\dot\varphi_1=-\alpha^2\varphi_1 + \alpha f_1^+\varphi_2 +
  f_2^+\varphi_1,\quad
  \dot\varphi_2=-\alpha^2\varphi_2 - \alpha f_1^+\varphi_1 +
  f_2^+\varphi_2,\\
 &2)&\dot\varphi_1=\alpha^2\varphi_1 + \alpha f_1^-\varphi_2 +
  f_2^-\varphi_1,\quad
  \dot\varphi_2=\alpha^2\varphi_2 + \alpha f_1^-\varphi_1 +
  f_2^-\varphi_2,
\end{eqnarray*}
where $f_i^{\pm} = f_i(\alpha^2(\varphi_2^2\pm \varphi_1^2))$.

\section{Conclusion}
In the papers \cite{FuZ2, FuZ3} we have constructed a number
of Ans\"atze (\ref{8}) which reduce the nonlinear heat equation
$u_t=[a(u)u_x]_x+f(u)$ to several ODEs. The basic technique
used was the anti-reduction method. The present paper provides
symmetry interpretation of the said results. It is important
to emphasize that there exist non-evolution equations which
also admit anti-reduction. In particular, in \cite{FuM, FuR, Zh}
an anti-reduction of the nonlinear acoustics equation, of
the equation for short waves in gas dynamics and of the nonlinear wave
equation is carried out. It would be of interest to extend the Theorem
1 in order to include into consideration these equations.

\section{Acknowledgments}
This work was supported by the Alexander von Humboldt Foundation.
Author would like to express his gratitude to Director of the Arnold
Sommerfeld Institute for Mathematical Physics Professor H.-D.Doebner
for invitation and hospitality.


\begin{thebibliography}{10}
\bibitem{BlC}
      Bluman G and Cole J 1969 {\it J. Math. Phys.} {\bf 18} 1025

\bibitem{ClKr}
      Clarkson P A and Kruskal M D 1989 {\it J. Math. Phys.} {\bf 30}
      2201

\bibitem{FuC} Fuchssteiner B and Carillo S 1990 {\it Physica A} {\bf
    166} 651

\bibitem{FuM}
      Fushchych W I and Mironyuk P I 1991{\it Proceedings of the
      Ukrainian Acad. Sci.} N6 23

\bibitem{FuN}
      Fushchych W I and Nikitin A G 1987 {\it Symmetries of Maxwell's
      Equations} (Dordrecht: Reidel)

\bibitem{FuR}
      Fushchych W I and Repeta V 1991 {\it Proceedings of the
      Ukrainian Acad. Sci.} N8 35

\bibitem{FuZ1}
      Fushchych W I and Zhdanov R Z 1992 {\it Ukrainian Math. J.}  {\bf
        44} 970

\bibitem{FuZ2}
      Fushchych W I and Zhdanov R Z 1993 {\it Proceedings of the
      Ukrainian Acad. Sci.} N11 37

\bibitem{FuZ3}
      Fushchych W I and Zhdanov R Z 1994 {\it J. Nonlinear Math. Phys.}
      {\bf 1} 60

\bibitem{Ga}
      Galaktionov V A 1990 {\it Diff. Int. Eq.} {\bf 3} 863

\bibitem{Ib}
      Ibragimov N Kh 1985 {\it Transformation Groups Applied to
        Mathematical Physics} (Boston: Reidel)

\bibitem{LeW}
      Levi D and Winternitz P 1989 {\it J. Phys. A} {\bf 22} 2915

\bibitem{OlR}
      Olver P J and Rosenau P 1987 {\it SIAM J. Appl. Math.} {\bf 18} 263

\bibitem{Ov}
      Ovsiannikov L V 1982 {\it Group Analysis of Differential Equations}
      (New York: Academic)

\bibitem{Po}
      Pommaret J F 1978 {\it Systems of Partial Differential Equations and
      Lie Pseudogroups} (New York: Gordon and Breach)

\bibitem{Sa} Samarskii A A, Galaktionov V A, Kurdyumov S P and
      Mikhailov A B 1994 {\it Blowing-up in problems for quasi-linear
        parabolic equations} (Berlin: De Gruyter)

\bibitem{Zh}
      Zhdanov R Z 1994 {\it J. Phys. A} {\bf 27} L291
\end{thebibliography}
\end{document}